\documentstyle[aps,prl,twocolumn,epsf]{revtex}
\draft
\begin{document}
\title{
Kondo Physics in the Single Electron Transistor with ac Driving}
\author{Peter Nordlander}
\address{
Department of Physics and Rice Quantum Institute,
Rice University, Houston, Texas 77251-1892
}
\author{Ned S. Wingreen}
\address{NEC Research Institute, 4 Independence Way,
Princeton, NJ 08540}
\author{Yigal Meir}
\address{Physics Department, Ben Gurion University, Beer
Sheva, 84105, Israel}
\author{David C. Langreth}
\address{Department of Physics and Astronomy,
Rutgers University, Piscataway, NJ 08854-8019
}

\maketitle
\thispagestyle{empty}
\begin{abstract}
\widetext
Using a time-dependent Anderson Hamiltonian,
a quantum dot with an ac voltage applied to a nearby gate
is investigated. A rich
dependence of the
linear response conductance on the external frequency and
driving amplitude is demonstrated.
At low frequencies the ac
potential produces sidebands of the Kondo peak in the spectral density
of the dot, resulting in a logarithmic decrease in
conductance over several decades of frequency.
At intermediate frequencies,
the conductance of the dot displays an oscillatory behavior
due to the appearance of Kondo resonances of the satellites
of the dot level.
At high frequencies, the conductance of the dot
can vary rapidly due to the interplay between photon-assisted tunneling
and the Kondo resonance.
\end{abstract}

\narrowtext

It has been predicted that, at low temperatures, transport through
a quantum dot should be governed by the same many-body phenomenon
that enhances the resistivity of a metal containing magnetic
impurities -- namely the Kondo effect \cite{glazman89}.
The recent observation of the Kondo effect by
Goldhaber {\it et al.}\cite{kastner97} in a quantum dot operating as a
single-electron
transistor  (SET) has fully verified these predictions. In contrast
to bulk metals, where the Kondo effect corresponds to the screening
of the free spins of a large number of magnetic impurities,
there is only one free spin in the
quantum-dot experiment. Moreover, a combination of bias and gate
voltages allow the Kondo regime, mixed-valence regime, and empty-site regime
all to be studied for the same quantum dot,
both in and out of equilibrium\cite{kastner97}.

Here we consider another opportunity presented by the
observation of the Kondo effect in a quantum dot that is not available
in bulk metals -- the application of an unscreened ac potential.
There is already a
large literature concerning the experimental application of
time-dependent fields to quantum dots \cite{acreview}.
For a dot acting as a Kondo system, the ac voltage can be used
to periodically modify the Kondo temperature or to alternate
between the Kondo and mixed-valence regimes. Thus
it is natural to ask %
what additional phenomena
occur in a driven system which
in steady state %
is dominated by the Kondo
effect \cite{hettler95,drivenkondo}.
Our results indicate a rich range of behavior with increasing
ac frequency,  from sidebands of the Kondo peak
at low ac frequencies, to conductance oscillations at intermediate
frequencies, and  finally to \pagebreak\vspace*{1.6in}
photon-assisted tunneling at high ac frequencies.

\begin{figure}[h]
\centerline{\epsfxsize=0.45\textwidth
\epsfbox{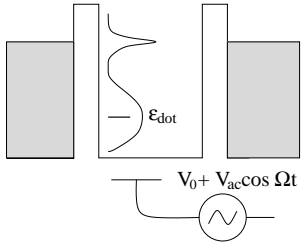}
}
\caption{Schematic picture of the quantum dot SET.
}
\label{fig:schematic}
\end{figure}

	The system of interest  is a semiconductor quantum dot, as  pictured
schematically in Fig.~\ref{fig:schematic}.
An electron
can be constrained
between two  reservoirs by
 tunneling barriers leading to a virtual electronic level within the dot at
energy
$\sim$$\epsilon_{\rm dot}$ (measured from the Fermi level) and width
$\sim$$2\Gamma_{\rm dot}$ \cite{Gnote}.
We assume that both the charging energy $e^2/C$ and the level spacing
in the dot are much larger than $\Gamma_{\rm dot}$, so the dot will operate
as a SET \cite{acreview}.
In this work, we consider only the
linear-response conductance between the two reservoirs. However, we will
allow an oscillating gate voltage
$V_{\rm g}(t)=V_0 +  V_{\rm ac}\cos\Omega t$ of arbitrary
(angular) frequency $\Omega$ and arbitrary
amplitude $V_{\rm ac}$, which modulates the virtual-level energy
$\epsilon_{\rm dot}(t)$.

	Such a system may be described by a constrained ($U=\infty$)
Anderson Hamiltonian
\begin{equation}
\sum_\sigma \!
\epsilon_{\rm dot}(t)\,
n_\sigma
+\sum_{k\sigma}\!\left[\epsilon_{k\sigma}n_{k\sigma}
+(V_k c^\dagger_{k\sigma}c_\sigma + {\rm H.c.})\right].
\label{hamiltonian}
\end{equation}
Here $c^\dagger_\sigma$ creates an electron of spin $\sigma$
in the quantum dot, while $n_\sigma$ is the corresponding
number operator; $c^\dagger_{k\sigma}$ creates a corresponding
reservoir electron; $k$ is shorthand for all other quantum numbers
of the reservoir electrons, including the designation of left or
right reservoir, while $V_k$ is the tunneling matrix element
through the appropriate barrier. Because the charging  energy to add
a second electron, $U = e^2/C$,
is assumed large, the Fock space in which the Hamiltonian
(\ref{hamiltonian}) operates is restricted to those
elements with zero or one electron in the dot.

At low temperatures,
the Anderson Hamiltonian (\ref{hamiltonian}) gives rise to the Kondo effect
when the level energy $\epsilon_{\rm dot}$ lies below the Fermi energy.
In this regime, a single electron occupies the dot which, in effect, turns
the dot into a magnetic impurity with a free spin. The temperature required to
observe the Kondo effect in linear response is of order
$T_{\rm K} \sim D \exp(-\pi|\epsilon_{\rm dot}|/\Gamma_{\rm dot})$,
where $D$ is the energy difference between the Fermi level and
the bottom of the band of states. For the
temperature range that is likely to be
experimentally accessible in a SET, $T \sim T_{\rm K}$
or higher, there exists a well tested and
reliable approximation known as
the non-crossing approximation (NCA)
\cite{bickers}. The NCA has been
formally generalized to the full time-dependent nonequilibrium case
\cite{LangrethNordlander91PRB},
and an exact method for the (numerical) solution implemented
\cite{ShaoetAl194PRB}. The time-dependent NCA
has been applied to Kondo physics in charge
transfer in hyperthermal ion scattering from metallic surfaces
\cite{ShaoetAl194PRB,chargetransfer} and
to energy transfer and stimulated desorption at metallic surfaces
\cite{energytransfer}.
 An independent formulation \cite{hettler95} has been
applied to quantum dots, although the high frequency expansion used there
appears limited. Here we present the exact time-dependent
NCA solution for a quantum dot over the full
range of applied frequencies.

	The time-dependent electronic structure of the dot can
be characterized by the time-dependent spectral density
\begin{equation}
\rho_{\rm dot}(\epsilon,t)\equiv \int_{-\infty}^{\infty}\frac{d\tau}{2\pi}
e^{i\epsilon\tau/\hbar} \langle
\{c_\sigma(t+\case{1}{2}\tau),c_\sigma^\dagger(t-\case{1}{2}\tau)\}\rangle
\end{equation}
 evaluated in the restricted Fock space.
For the equilibrium Kondo system,
$\rho_{\rm dot}(\epsilon)$
is time independent, and looks like the graph in the schematic in
Fig.~\ref{fig:schematic}. Roughly speaking, $\rho_{\rm dot}(\epsilon)$
consists of a
broad peak of width $\sim$$2\Gamma_{\rm dot}$ at the level position
$\epsilon_{\rm dot}$ and a sharp Kondo peak of width
$\sim$$T_{\rm K}$ near the Fermi level.  We will refer to these
features
as the virtual-level peak and the Kondo peak, respectively.
In the steady-state case, the linear-response 
conductance $G$ through a  dot
symmetrically coupled to two reservoirs
is given by \cite{meir92}
\begin{equation}
G = {e^2 \over \hbar} {{\Gamma_{\rm dot} } \over {2} }
\int \!d\epsilon\,\rho_{\rm dot} (\epsilon)\left(-\frac{\partial
f(\epsilon)}{\partial\epsilon}\right),
\label{conductance}
\end{equation}
where
$f(\epsilon)$ is the Fermi function.  The formula
(\ref{conductance}) will still be valid in the case where the
gate voltage is time dependent if $G$ is the {\it time-averaged} conductance
and $\rho_{\rm dot}(\epsilon)$ is replaced by the \nopagebreak {\it
time-averaged} spectral density
$\langle\rho_{\rm dot}(\epsilon,t)\rangle$
\cite{jauho94,rhonote}.  For a given system, this average will depend on the
driving amplitude $V_{\rm ac}$ and frequency $\Omega$.

\begin{figure}[t]
\centerline{\epsfxsize=0.45\textwidth
\epsfbox{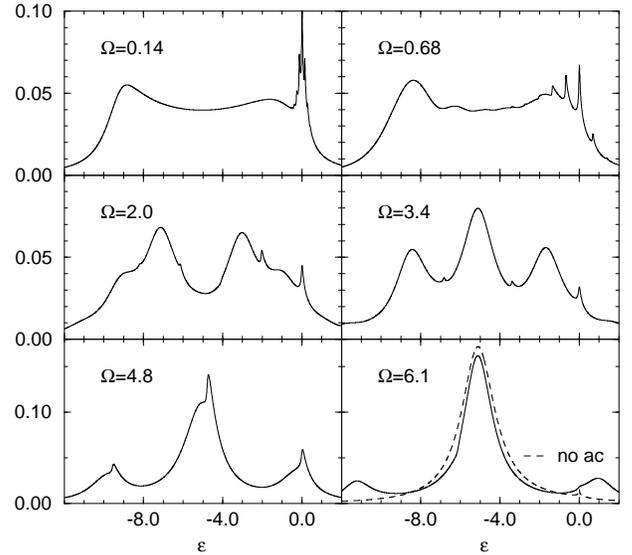}
}
\vspace*{.2cm}
\caption{The spectral density $\langle\rho_{\rm dot}(\epsilon,t)\rangle$
vs. energy $\epsilon$
for a quantum dot with level energy $\epsilon_{\rm dot}(t)=-5 +4 \cos \Omega t$
and $ T = 0.005$.
The non-driven case is also shown in the final panel.
Throughout this letter, energies are
in units of   $\Gamma_{\rm dot}$.
}
\label{fig:spectral}
\end{figure}

\begin{figure}[h]
\centerline{\epsfxsize=0.45\textwidth
\epsfbox{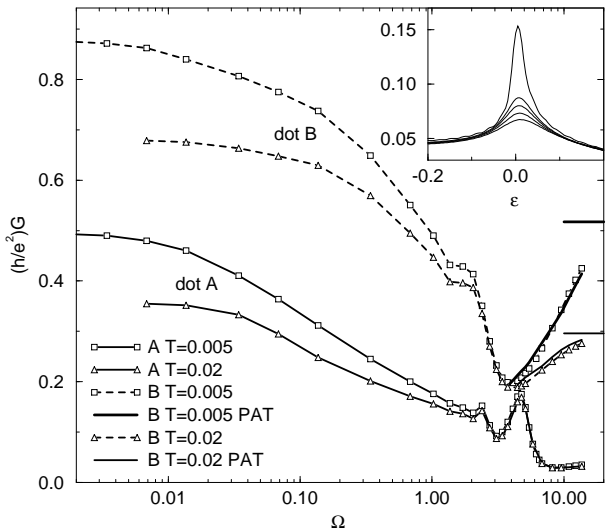}
}
\vspace*{.2cm}
\caption{Conductance of two different quantum dots, each at two
different temperatures: Dot A, $\epsilon_{\rm dot}(t) = -5 + 4 \cos \Omega t$;
Dot B, $\epsilon_{\rm dot}(t) = -2.5 + 2 \cos \Omega t$.
The curves  at the high $\Omega$ end for
dot B (marked ``PAT'') are from our photon-assisted-tunneling model,
while the exact high frequency asymptotes for dot B
are shown as short horizontal lines extending from
the right vertical axis. The inset shows the spectral density
$\langle \rho_{\rm dot} (\epsilon, t)\rangle$ of dot
B around the Fermi level, at $ T = 0.005$,
for large frequencies,
from $\Omega=4.8$ (lowest curve), through 5.5, 6.1, 6.8 to
$\Omega=14$ (topmost curve).
}
\label{fig:conductance}
\end{figure}

In Fig.~\ref{fig:spectral} we show the calculated
$\langle \rho_{\rm dot}(\epsilon,t)\rangle$
as a function of energy $\epsilon$ for a level with energy
$\epsilon_{\rm dot}(t) = \epsilon_{\rm dot} + \epsilon_{\rm ac} \cos
\Omega t$ at several different
frequencies $\Omega$.  The corresponding conductance is
shown by the curve labeled dot A, $T=0.005$ in
Fig.~\ref{fig:conductance}.
For the lowest $\Omega$, the response of
the system is relatively adiabatic and the displayed spectral
function resembles the spectral function that would have
resulted if the system had been in perfect equilibrium for all the
dot level positions over a period of oscillation of $\epsilon_{\rm dot}(t)$.
The two broad peaks are the influence of the virtual level peaks
at the two stationary points of this oscillation (here at $\epsilon=-1$
and $\epsilon=-9$).
As the frequency $\Omega$ is increased, marked nonadiabatic effects
result, the most obvious being the appearance of multiple satellites
around the Kondo resonance \cite{hettler95}.
 These sidebands appear at energies
equal to $\hbar$ times multiples of the driving frequency $\Omega$
\cite{diff_cond_peaks}.
As the
frequency $\Omega$ is increased, spectral weight is transferred
from the main Kondo peak  to these satellites.
 As the conductance is
dominated by $\langle \rho_{\rm dot}(\epsilon,t)
\rangle$ at the Fermi energy, this causes the slow
logarithmic falloff of the conductance over two decades of
frequency, as shown in Fig.~\ref{fig:conductance}.

As $\hbar\Omega$ becomes larger than $\Gamma_{\rm dot}$, inspection of
Fig.~\ref{fig:spectral} shows that broad
satellites also appear  at energy separations
$n \hbar \Omega$ around the average
 virtual-level position $\epsilon_{\rm dot}$.
These satellites of the virtual level are the analogues of those predicted in
the noninteracting case \cite{jauho94},  which decrease in magnitude as
the order $n$ of the Bessel function
$J_n$. Here, however, the virtual-level satellites have their own Kondo peaks;
each of the latter gets strong when the corresponding virtual-level satellite
reaches a position  a little
below the Fermi level, and then disappears as the broad satellite
crosses the  Fermi level. This effect produces the
oscillations in the conductance that are evident in the
lower curves in Fig.~\ref{fig:conductance}.
These oscillations are  very different from those that would occur in a
noninteracting ($U=0$) case:  due to the Kondo peaks they are substantially
stronger, their maxima occur at different frequencies
and their magnitudes are temperature dependent.
As the last virtual-level satellite crosses the Fermi level,
$\hbar \Omega = |\epsilon_{\rm dot}|$, the
dot level energy begins to vary too fast for the system to respond and
the average spectral function approaches (exactly
as $\Omega\rightarrow\infty$)  the {\it equilibrium}
spectral function for a dot level centered at the average
position $\epsilon_{\rm dot}$.
For the parameters of Fig.~\ref{fig:spectral}, the high frequency
 region is uninteresting, because the temperature is far above
the Kondo temperature. Therefore, the conductance shows little
temperature or frequency dependence at these high frequencies.

The situation is quite different for the system (dot B) displayed in the
upper two curves in Fig.~\ref{fig:conductance}, which
 displays a strong
Kondo effect when the dot level is held at its average energy
$\epsilon_{\rm dot}$.
In this case the $\Omega \rightarrow \infty$
 conductance is strongly enhanced by
the Kondo effect, and is consequently temperature dependent as well.
Note that the conductance falls off significantly from its asymptotic,
$\Omega \rightarrow \infty$, value for frequencies still much larger than
either the depth of the level $|\epsilon_{\rm dot}|$ or its %
width.
This effect is due to a rapid decline of the amplitude of the
Kondo peak in the spectral density, as illustrated
in the inset of Fig.~\ref{fig:conductance}. We
propose the following explanation for this phenomenon. The energy
$\hbar\Omega$ excites the dot, producing satellites
\cite{TienGordon,jauho94}
of the virtual level peak at energies $\epsilon_{\rm dot} \pm n \hbar \Omega$,
 which, for $\hbar\Omega \gg \Gamma_{\rm dot}$
 have strength roughly
given by $[J_n(\epsilon_{\rm ac}/\hbar\Omega)]^2$ as
in the $U=0$ case (see Fig.~\ref{fig:spectral} and the previous discussion).
For large $\hbar\Omega$, only the two $n=1$ satellites
have any significant strength, and the higher
 lies above the Fermi level, allowing
an electron on the dot to decay
at the rate
$(1/\hbar) \Gamma_{\rm dot}(\epsilon_{\rm dot}+\hbar\Omega)$.
The overall electron decay
probability per unit time $\Gamma_{\rm decay}/\hbar$ due to this
photon-assisted-tunneling mechanism (PAT) is therefore given by
\begin{equation}
\Gamma_{\rm decay} \approx
	[J_1(\epsilon_{\rm ac}/\hbar\Omega)]^2
	\,\Gamma_{\rm dot}(\epsilon_{\rm dot}+\hbar\Omega).
\label{gammadecay}
\end{equation}
The above rate carries with it an energy uncertainty, which we speculate
has roughly the same effect on the Kondo peak as the energy smearing
due to a finite temperature. We can test this conjecture by
calculating the {\it equilibrium} conductance at an {\it effective}
temperature $T_{\rm eff}$ given by
$ T_{\rm eff} =  T + \Gamma_{\rm decay}$. The results of such
a calculation are shown in Fig.~\ref{fig:conductance} (PAT curves),
 where they compare
very favorably with our results for the  conductance in the ac-driven
system.

Returning to the behavior at low frequencies, we find that
it can be best understood in terms of
the Kondo Hamiltonian, which, with respect to properties
near the Fermi level,  is equivalent to
the Anderson Hamiltonian (\ref{hamiltonian})
in the extreme Kondo region
 $-\epsilon_{\rm dot} \gg \Gamma_{\rm dot}$ \cite{SchriefferWolff}.
In this limit the dot can be replaced simply by a dynamical  Heisenberg
spin $\vec{S}$ ($S^2=\case{3}{4}$), which
scatters electrons both within and between reservoirs.
The Kondo Hamiltonian corresponding to the  Anderson model
(\ref{hamiltonian})  is
\begin{equation}
\sum_{kk'\sigma\sigma'}J_{kk'}(t)
\left(\vec{S}\cdot\vec{\sigma}_{\sigma\sigma'}+
\case{1}{2}\delta_{\sigma\sigma'}\right)
c^\dagger_{k\sigma}c_{k'\sigma'},
\label{hamsd}
\end{equation}
where the components of $\vec{\sigma}$ are the Pauli spin matrices.
For near Fermi level properties we can suppress the detailed $k$ dependence
of $J$ and $V$ and introduce a large energy cutoff $D$ \cite{Dnote}, in which
case the relationship between the Kondo and Anderson Hamiltonians
\cite{SchriefferWolff}
is $J(t)=|V^2/\epsilon_{\rm dot}(t)|$ for our $U = \infty$ case.
 If we let $w_{\rm leads}(\epsilon)/\hbar$ be the total
rate at which lead electrons of energy $\epsilon$
undergo intralead and interlead scattering
by the dot, then $w_{\rm leads}(\epsilon)$
will have a Kondo peak for $\epsilon$ near the Fermi level. Furthermore,
if $J$ is modulated as
$J(t)= {\langle{J}\rangle}(1+\alpha \cos \Omega t)$, then
an electron scattered by the dot will be able to absorb or emit
multiple quanta of energy $\hbar \Omega$, leading to
satellites of the Kondo peak in
$\langle\rho_{\rm dot}(\epsilon,t)\rangle$
through the exact Anderson model relation
\begin{figure}[h]
\centerline{\epsfxsize=0.45\textwidth
\epsfbox{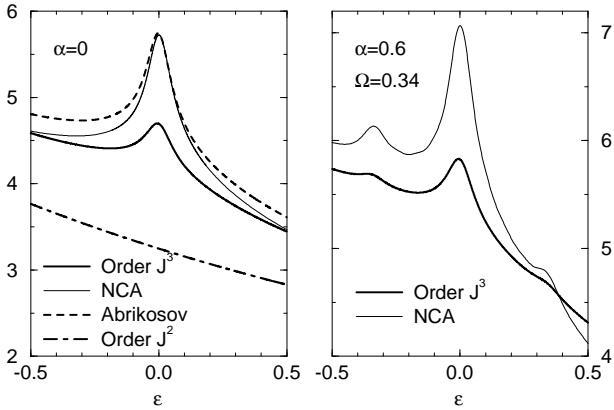}
}
\vspace*{0.2cm}
\caption{Spectral density
$\langle\rho_{\rm dot}(\epsilon,t)\rangle$ times $10^3$
in the Kondo model and NCA for $ T = 0.02$
and ${\langle{J}\rangle}\rho=0.023$ ($\epsilon_{\rm dot} = -7 $).
For the  non-driven case (left panel) we
also show the comparable result from summing all the leading logarithmic
terms (Abrikosov, Ref.~\protect\onlinecite{abrikosov}), as well as that
obtained to order $J^2$.
The energy dependence of ${\langle{J}\rangle}$
(Ref.~\protect\onlinecite{SchriefferWolff})
has been included to order $J^2$ in all the Kondo Hamiltonian curves.
}
\label{fig:J3}
\end{figure}
\noindent $w_{\rm leads}(\epsilon)$. This
then reflects back on  %
\begin{equation}
w_{\rm leads}(\epsilon) = \Gamma_{\rm dot}(\epsilon)
\langle\rho_{\rm dot}(\epsilon,t)\rangle / \rho_{\rm leads}(\epsilon),
\label{wleads}
\end{equation}
where $\rho_{\rm leads}(\epsilon)$
is the state density per spin in the leads.

The  above can be illustrated explicitly using perturbation
theory in $J$. Keeping all terms of order $J^2$ and logarithmic terms
to order $J^3$, we find, using a nonequilibrium version of
 Abrikosov's pseudofermion technique
\cite{abrikosov}, that
\begin{equation}
w_{\rm leads}(\epsilon) =
2\pi \langle J^2\rangle \rho\bigg[ 1+3\langle{J}\rangle\rho\sum_{n=-1}^1
a_n g(\epsilon+n\hbar\Omega)\bigg],
\label{GammaLeads}
\end{equation}
where $\rho=\rho_{\rm leads}(0)$,
$a_0=1$, $a_{\pm 1} = \alpha^2/(2+\alpha^2)$,
$\langle J^2\rangle=(1+\case{1}{2}\alpha^2){\langle{J}\rangle}^2$,
and
\begin{equation}
g(\epsilon)= \frac{1}{2}\int_{-D}^{D}\!d\epsilon'\,
	\frac{1-2f(\epsilon')}{\epsilon'-\epsilon}
\rightarrow \ln\left|\frac{D}{\epsilon}\right|,
\label{gdef}
\end{equation}
the last limit being approached when $ T \ll |\epsilon|$. In
Fig.~\ref{fig:J3} we compare this prediction with the full
NCA theory. Although we are not strictly in the parameter region
where the $J^3$ theory is quantitatively valid, the qualitative
agreement is quite satisfactory.

The present results indicate rich behavior when an external ac potential
is applied to a quantum dot in the regime where the
conductance is dominated by the Kondo effect. While the time-dependent
NCA method  employed spans the full range of applied frequency,
some additional insight has been gained into the behavior both at very
low and very high frequencies. At low frequencies a time-dependent Kondo model
helps explain the amplitudes of sidebands of the Kondo peak in the spectral
density of the dot. At high frequencies, a cutoff
of the Kondo peak due to photon-assisted tunneling processes accounts for
the reduction of conductance. We hope that our work will inspire
experimental investigation of these phenomena and other ramifications
of ac driving applied to Kondo systems.

The work
was supported in part by NSF grants DMR 95-21444 (Rice) and
 DMR 97-08499 (Rutgers), and by
  US-Israeli Binational Science
Foundation grant 94-00277/1 (BGU).


\begin{thebibliography}{10}

\bibitem{glazman89} L.~I. Glazman and M.~E. Raikh, Pis'ma Zh. Eksp. Teor.
Fiz. {\bf 47}, 378 (1988) [JETP Lett. {\bf 47}, 452 (1988)]; T.~K. Ng
and P.~A. Lee, Phys. Rev. Lett. {\bf 61}, 1768 (1988);
S. Hershfield, J.~H. Davies, and J.~W. Wilkins, Phys. Rev. Lett.
{\bf 67}, 3720 (1991);
Y. Meir, N.~S. Wingreen, and P.~A. Lee, Phys. Rev. Lett.
{\bf 70},  2601 (1993);
N.~S. Wingreen and Y. Meir, Phys. Rev. B {\bf 49},  11\,040  (1994).

\bibitem{kastner97}
D. Goldhaber-Gordon {\it et~al.}, Nature {\bf 391}, 156 (1998).

\bibitem{acreview} L.~P. Kouwenhoven {\it et al.}, in {\it
Mesoscopic Electron Transport}, edited by L.~L. Sohn, L.~P. Kouwenhoven,
and G. Sch\"on (Kluwer, Netherlands, 1997).

\bibitem{hettler95}
M.~H. Hettler and H. Schoeller, Phys. Rev. Lett. {\bf 74},  4907  (1995).

%



%
%

\bibitem{drivenkondo}
 A. Schiller and S. Hershfield, Phys. Rev. Lett.
     {\bf 77}, 1821 (1996);
T.~K. Ng, Phys. Rev. Lett. {\bf 76},  487  (1996);
Y.~Goldin and Y.~Avishai, preprint cond-mat/9710085.



\bibitem{Gnote}
%
 %
%
We define $\Gamma_{\rm
  dot}(\epsilon) = 2
  \pi \sum_k |V_k|^2 \delta(\epsilon -\epsilon_k)$, a slowly varying quantity.
 The
  notation $\Gamma_{\rm dot}$ with no energy specified will always refer the
  value at the Fermi level.

%

\bibitem{bickers}
N. E. Bickers, Rev.\ Mod.\ Phys.\ {\bf 59}, 845 (1987).

%
%

%
%
%

%
%
%

\bibitem{LangrethNordlander91PRB}
D.~C. Langreth and P. Nordlander, Phys. Rev. B {\bf 43},  2541  (1991).

%
%


 \bibitem{ShaoetAl194PRB}
 H. Shao, D.~C. Langreth, and P. Nordlander, Phys. Rev. B {\bf 49},  13\,929
   (1994).

\bibitem{chargetransfer}
 H. Shao, P. Nordlander, and D.~C. Langreth, Phys. Rev. B
 {\bf 52},  2988  (1995), Phys. Rev. Lett. {\bf 77},  948   (1996).

%
%
%
%
%
%
%
%
\bibitem{energytransfer}
  T. Brunner and D.~C. Langreth, Phys. Rev. B {\bf 55},  2578  (1997);
M. Plihal and D.~C. Langreth, Surf. Sci. Lett. {\bf 395}, 252 (1998);
Phys. Rev. B (submitted).

%
%
%
%
%
%
%
%
%

%
%
%

\bibitem{meir92}
Y. Meir and N.~S. Wingreen, Phys. Rev. Lett. {\bf 68},  2512  (1992).

\bibitem{jauho94}
A.-P. Jauho, N.~S. Wingreen, and Y. Meir, Phys. Rev. B {\bf 50},  5528  (1994).

\bibitem{rhonote} For the Hamiltonian (\ref{hamiltonian}) the time average
$\langle
\rho_{\rm dot}(\epsilon,t)\rangle\equiv-{\rm Im}\langle
A(\epsilon,t)\rangle/\pi$,
where $A(\epsilon,t)$ is the  retarded and hence causal function defined in
Ref.~\onlinecite{jauho94}, Eq.~(28).


%
%
%
%
%
%

\bibitem{diff_cond_peaks}
Hence in an experiment one 
may also expect peaks spaced by $\hbar \Omega$ in the
differential conductance.
 

\bibitem{TienGordon}
P.~K. Tien and J.~P. Gordon, Phys. Rev. {\bf 129},  647  (1963).

\bibitem{SchriefferWolff}
J.~R. Schrieffer and P.~A. Wolff, Phys. Rev. {\bf 149},  491  (1966).

\bibitem{Dnote}
For the lead state density used in the NCA calculations, $\rho_{\rm
  leads}(\epsilon)= \rho_{\rm leads}(0)[1 - \epsilon^2/(20\Gamma_{\rm dot}
)^2]$,
  the appropriate value is given by $D=20\Gamma_{\rm dot}/\sqrt{e}$; see
  Ref.~\onlinecite{ShaoetAl194PRB}.

\bibitem{abrikosov}
A.~A. Abrikosov, Physics {\bf 2},  5  (1965).

\end{thebibliography}
\end{document}